\documentclass[preprint,showpacs,preprintnumbers,amsmath,amssymb,prb]{revtex4}

\usepackage{amsmath}
\usepackage{amssymb}
\usepackage{graphicx} 
\usepackage{color}
\usepackage{dcolumn}
\usepackage{bm}
\usepackage{enumerate}

\newcommand{\dps}[1]  {\displaystyle{#1} }

\begin{document}

\preprint{APS/}

\title{Computation of free energy profiles with parallel adaptive dynamics}

\author{Tony Leli\`evre}
\email{lelievre@cermics.enpc.fr}
\altaffiliation[Also at ]{INRIA Rocquencourt, MICMAC team, B.P. 105, 78153 Le Chesnay Cedex, France.}
\author{Mathias Rousset}
\email{rousset@cermics.enpc.fr}
\altaffiliation[Also at ]{INRIA Rocquencourt, MICMAC team, B.P. 105, 78153 Le Chesnay Cedex, France.}
\author{Gabriel Stoltz}
\email{stoltz@cermics.enpc.fr}
\altaffiliation[Also at ]{CEA/DAM Ile-de-France,
BP 12, 91680 Bruy\`eres-le-Ch\^atel, France.}
 \affiliation{%
 CERMICS, Ecole Nationale des
  Ponts et Chauss\'ees (ParisTech), 6 \& 8 Avenue Blaise Pascal,
  77455 Marne-la-Vall\'ee, France.
}%

\date{\today}

\begin{abstract}
We propose a formulation of adaptive computation of free energy differences, in the ABF or nonequilibrium metadynamics spirit, using conditional distributions of samples of configurations which evolve in time. This allows to present a truly unifying framework for these methods, and to prove convergence results for certain classes of algorithms. From a numerical viewpoint, a parallel implementation of these methods is very natural, the replicas interacting through the reconstructed free energy. We show how to improve this parallel implementation by resorting to some selection mechanism on the replicas. This is illustrated by computations on a model system of conformational changes.
\end{abstract}


\maketitle


\section{Introduction}

One of the most important goals of molecular simulation is the
computation of free energy differences as a function of some selected
degrees of freedom, called reaction coordinates. The dynamics of the
system can indeed often be split into slowly evolving degrees of
freedom, which determine the reaction coordinates to be used, and other
rapidly evolving degrees of freedom. The free energy differences allow
to characterize global changes in the system under study, 
and give
information about the relative stabilities of several species, as well as their transition kinetics.
However, the
free energy barriers to overcome are so large in many applications that
a computation based on a straightforward sampling is unfeasible since the system remains stuck in metastable free energy sets.

A classical technique to compute free energy differences is thermodynamic integration, dating back to Kirkwood~\cite{Kirk35}, which mimics the
quasi-static evolution of a system as a succession of equilibrium
samplings, which amounts to an infinitely slow switching between the
initial and final states. Another classical technique is the free energy perturbation
method, introduced by Zwanzig~\cite{Zwanz54}, which recasts free energy
differences as a phase-space integral, so that usual sampling techniques
can be employed. Notice also that there exist many refinements for those
two classes of techniques, such as umbrella sampling~\cite{TV77}. 

More recently, methods relying on nonequilibrium dynamics have emerged. They follow the pioneering work of
Jarzynski~\cite{Jarzynski97}, or use some adaptive dynamics such as
the Wang-Landau approach~\cite{WL01}, the adaptive biasing force
(ABF)~\cite{DP01,DPW02,HC04}, or the nonequilibrium
metadynamics~\cite{BLP06}. These approaches use the whole history
of the exploration process to bias the current dynamics in order to
force the escape from metastable sets. 
This is done by simultaneously estimating the free energy from an evolving ensemble of configurations of the dynamics, and using this estimate to bias the dynamics, so that the effective free energy surface explored is flattened. In the long time limit, the bias exactly gives the actual free energy profile.
Adaptive methods could therefore be seen as umbrella sampling with an evolving potential. This was already noticed in a previous study presenting an adaptive dynamics as a 'self-healing umbrella sampling'~\cite{MBCPS06}.

To present the adaptive methods mentioned above in a general and unifying framework, it is convenient, as is done in~\cite{BLP06}, to consider ensemble of realizations (see Eq.~(\ref{eq:straightforward_parallel})). The system is then described by the distribution of the configurations of this ensemble in the limit of an infinite number of replicas simulated in parallel. The key point is to reformulate the computation of the bias of adaptive dynamics, using conditional distributions (that is, distribution of the configurations for a given value of the reaction coordinate) of the latter sample. This was already proposed in~\cite{EV} in the equilibrium case, and is somewhat implicit in~\cite{BLP06}. 
This concept clarifies the presentation of adaptive methods, allows mathematical proofs of
convergence~\cite{LORS} or at least, existence of a stationary state of the dynamics (still in the case of an infinite number of replicas), and suggests natural numerical strategies: the discretization may be done through a parallel implementation of
several replicas of the system, which all contribute to construct the
free energy profile. Such a parallel implementation was already proposed
in~\cite{RLLgMP06} in the case of metadynamics. We show here how an
additional selection process on the replicas can enhance the sampling of
the reaction coordinates in comparison with a straightforward parallel
implementation.

The paper is organized as follows. In Section~\ref{sec:general_framework}, we describe the general formalism for adaptive dynamics, using conditional probabilities, and show how to update the biasing potential in order to compute the free energy profile in the longtime limit, using a fixed-point strategy. Some applications of this formalism are presented in Section~\ref{sec:application_dynamics}, and allow to recover the usual adaptive dynamics such as the nonequilibrium metadynamics, the Wang-Landau scheme or the ABF method. We then discuss possible parallel implementation strategies in Section~\ref{sec:implementation}. In particular, it is shown in Section~\ref{sec:selection} how a selection process can enhance the straightforward parallel implementation. This is finally illustrated by numerical results for a toy model of conformational changes in Section~\ref{sec:numerical}.


\section{A general framework for adaptive dynamics}
\label{sec:general_framework}

\subsection{Notations}

For a system described by a potential $V(q)$, the Boltzmann measure 
in the canonical ensemble is $Z^{-1} \exp\left(-\beta V(q) \right) \, dq $ (where $Z$ is a
normalization constant, the so-called partition function). 
Consider a reaction coordinate $\xi$, taking values in the one
dimensional torus, or in the interval $[0,1]$. In the latter case, reflecting boundary conditions for the dynamics on
the two extremal values $\xi(q)=0$, $\xi(q)=1$ are used.
The free
energy (or potential of mean force (PMF)) to be computed is defined up to an additive constant by the normalization of a Boltzmann average of the configurations restricted to a given value of the reaction coordinate:
\begin{equation}
\label{eq:PMF}
A(z) = - \beta^{-1} \ln \int \exp(-\beta V(q) ) \, \delta_{\xi(q)-z}.
\end{equation}
The PMF $A$ exactly gives the Boltzmann weights of the equilibrium
distribution of the reaction coordinate. The
so-called {\it mean force} $A'(z)$ along the reaction coordinate can also be
expressed as the Boltzmann average of a real-valued ``force'' field
$f^V$ over the configurations restricted to a given value of the reaction coordinate $\xi(q)=z$:
\begin{equation}
\label{eq:biasing_force}
A'(z)  = \frac{\dps \int f^V(q) \, \exp(-\beta V(q) ) \, \delta_{\xi(q)-z}}{\dps \int \exp(-\beta V(q) ) \, \delta_{\xi(q)-z}}.
\end{equation}
Here and in the sequel, we denote by $A'$ the derivative of $A$ with respect to $z$. 

The force field $f^V$ can be expressed only in terms of derivatives of first and second order of the reaction coordinate $\xi$ as
\begin{equation}
\label{eq:local_force}
f^V= \frac{\nabla V \cdot \nabla \xi}{|\nabla \xi|^2} - \beta^{-1} {\rm div} \left(\frac{\nabla
    \xi}{|\nabla \xi|^2} \right).
\end{equation}
For the mathematical derivation of this formula, and extension to
multi-dimensional reaction coordinates or phase-spaces with holonomic
constraints, we refer for example to~\cite{OB98,CKV05,CLV06}.

\subsection{Adaptive biasing dynamics}

Adaptive dynamics are defined through the dynamics used, which dictates the distribution of the configurations at equilibrium (see Section~\ref{sec:dynamics}), a biasing potential (Section~\ref{sec:biasing_pot}), and the way this potential is updated (see Section~\ref{sec:update_eq} for a heuristic derivation in the equilibrium case motivating the general setting of Section~\ref{ref:update_noneq}). 

\subsubsection{The dynamics}
\label{sec:dynamics}

Trajectories $t \mapsto Q_t$ are computed according to some dynamics which are ergodic with respect to the Boltzmann measure when the potential is time-independent. 
For instance, the Langevin dynamics or the overdamped Langevin dynamics
may be used. We will denote by~$\psi_t(q)$ the probability distribution (or density) of configurations at time~$t$. This distribution will be used to update the biasing potential~$A_{\rm bias}$. 

From a practical point of view, when $M$ replicas $(Q^{i,M}_t)_{i=1,\dots,M}$ of the system are simulated in parallel, the density of states $\psi_t(q)$ is approximated by the instantaneous distribution of the replicas
\begin{equation}
\label{eq:straightforward_parallel}
\psi_t(q) = \lim_{M \to +\infty} \frac1M \sum_{i=1}^M \delta_{Q^{i,M}_t - q}.
\end{equation}
In some cases, the density of states can also be approximated using the distribution of configurations along the trajectory, relying on some ergodic assumption.

The definition of adaptive methods requires the definition of two important quantities obtained from the distribution $\psi_t(q)$. The first one is the distribution $\psi^\xi_t$ of the reaction coordinate values, which is the marginal law of $\psi_t$ with respect to $\xi$:
\begin{equation}
\label{eq:marginal_law}
\psi^\xi_t(z) = \int \psi_t(q) \, \delta_{\xi(q) - z}.
\end{equation}
This quantity will be useful to propose a biasing potential (see Eqs.~(\ref{eq:ABP_calcul})-(\ref{eq:ABP_update})). Another important
quantity is the conditional average of some function $h$ for some fixed value of the reaction coordinate:
\begin{equation}
\label{eq:conditional_average}
\langle h \rangle_{t,z} = \frac{\int h(q) \psi_t(q) \, \delta_{\xi(q)-z}}{\int \psi_t(q) \, \delta_{\xi(q)-z}}. 
\end{equation}
Such averages are used to propose biasing forces (see Eqs.~(\ref{eq:ABF_calcul})-(\ref{eq:ABF_update})).


\subsubsection{The biasing potential}
\label{sec:biasing_pot}

In adaptive dynamics, the interaction potential is time-dependent:
\begin{equation}
\label{eq:TDpotential}
{\cal V}_t(q) = V(q)-A_{\text{bias}}(t,\xi(q)).
\end{equation}
The biasing potential $A_{\rm bias}$, whose precise form varies according
to the method under study, depends only on $q$ through the reaction coordinate value $\xi(q)$ and is updated using the history of the configurations. It is expected that this biasing potential converges (up to an additive constant) toward the free energy $A$ given by~(\ref{eq:PMF}) in the long-time limit, so that the equilibrium distribution of the reaction coordinate is the uniform distribution.

The key idea common to all adaptive methods is to resort to a fixed point strategy, in order for the observed free energy to converge to a constant or the mean force to vanish, and the dynamics to reach equilibrium (see the updates~(\ref{eq:eq_ABP}) or~(\ref{eq:eq_ABF}) in the equilibrium case and~(\ref{eq:ABP_update}) or~(\ref{eq:ABF_update}) in the nonequilibrium case).

\subsubsection{Updating the biasing potential - The equilibrium case}
\label{sec:update_eq}

To derive a possible form for the biasing potential, let us first assume that the system is instantaneously at equilibrium 
with respect to the biased potential ${\cal V}_t$, {\it i.e.} $Q_t$ has density 
$\psi^{\rm eq}_t(q) = Z_t^{-1} \exp(-\beta {\cal V}_t(q))$. 
In this case, resorting to~(\ref{eq:PMF}), the {\em observed free energy} (see~(\ref{eq:ABP_calcul}) for a general definition) is
\begin{equation}
\label{eq:calcul_ABP_eq}
-\beta^{-1} \ln \int \psi^{\rm eq}_t(q) \, \delta_{\xi(q)-z} = A(z) - A_{\rm bias}(t,z) + \beta^{-1} \ln Z_t. 
\end{equation}
Thus, for a characteristic time $\tau$ to be chosen, an update of $A_{\rm bias}$ of the form
\begin{equation}
\label{eq:eq_ABP}
\partial_t A_{\rm bias}(t,z) = -\frac{\beta^{-1}}{\tau} \ln \int \psi^{\rm eq}_t(q) \, \delta_{\xi(q)-z}
\end{equation}
is such that $A'_{\rm bias}(t) \to A'$ when $t \to +\infty$ exponentially fast with rate $1/\tau$. Notice that we stated the convergence in terms of the mean force, because, in view of the constant term~$\beta^{-1} \ln Z_t$ in Eq.~(\ref{eq:calcul_ABP_eq}), the potential of mean force only converges up to a constant to the true potential of mean force.

Similar considerations
hold for the mean force: replacing the potential $V$ with ${\cal V}_t$
given by~(\ref{eq:TDpotential}), and resorting to~(\ref{eq:biasing_force})-(\ref{eq:local_force}), the observed mean force (see~(\ref{eq:ABF_calcul}) for a general definition) is
\begin{equation}
\label{eq:calcul_ABF_eq}
\frac{\int f^{{\cal V}_t}(q) \, \psi^{\rm eq}_t(q) \, \delta_{\xi(q)-z}}{\int \psi^{\rm eq}_t(q) \, \delta_{\xi(q)-z}} = A'(z) - A'_{\rm bias}(t,z),
\end{equation}
since $f^{{\cal V}_t}(q) = f^V(q) - A'_{\rm bias}(t,\xi(q))$.
An update of $A'_{\rm bias}(t)$ of the form
\begin{equation}
\label{eq:eq_ABF}
\partial_t A'_{\rm bias}(t,z) = \frac{1}{\tau} \, \frac{\int f^{{\cal V}_t}(q) \, \psi^{\rm eq}_t(q) \, \delta_{\xi(q)-z}(dq)}{\int \psi^{\rm eq}_t(q) \, \delta_{\xi(q)-z}(dq)}
\end{equation}
is therefore such that $A'_{\rm bias}(t) \to A'$ when $t \to +\infty$ exponentially fast with rate $1/\tau$.

\subsubsection{Updating the biasing potential - The nonequililibrium case}
\label{ref:update_noneq}

Now, in general, the system is not at equilibrium for the potential ${\cal V}_t$: $\psi_t \not = \psi_t^{\rm eq}$. We use the above procedure as a guideline to update the biasing potential $A_{\text{bias}}(t,z)$. To derive equations for the biasing potential, let us first define two quantities. The first one is the {\em observed free energy} or the {\em observed potential of mean force}, defined as
\begin{equation}
\label{eq:ABP_calcul}
A_{{\rm pot}, \text{obs}}(t,z)=-\beta^{-1} \ln \int \psi_t(q) \, \delta_{\xi(q)-z}.
\end{equation}
This quantity can be interpreted as the free energy associated with the ensemble of configurations with density of states $\psi_t(q)$ (see Eq.~(\ref{eq:PMF})). The observed free energy $A_{{\rm pot}, \text{obs}}(t,z)$ is high when the number of visited states with reaction coordinate value $z$ is small. In the long-time limit, the distribution of the reaction coordinate is expected to be uniform, so that the observed free energy is constant.

In the same way, the {\em observed mean force} is defined as the conditional average of the time-dependent biasing force for a given value of the reaction coordinate:
\begin{eqnarray}
\label{eq:ABF_calcul}
A_{{\rm force}, \text{obs}}'(t,z) = \frac{\int f^{{\cal V}_t}(q) \, \psi_t(q) \, \delta_{\xi(q)-z}}{\int \psi_t(q) \, \delta_{\xi(q)-z}}
= \frac{ \int f^{V}(q) \, \psi_t(q) \, \delta_{\xi(q)-z}}{\int \psi_t(q) \, \delta_{\xi(q)-z}}-A'_{\text{bias}}(t,z).
\end{eqnarray}
This quantity can be interpreted as the mean force associated with $\psi_t(q)$ (see Eqs.~(\ref{eq:biasing_force})-(\ref{eq:local_force})), minus the biasing force at time $t$. It is expected to vanish in the long-time limit, so that the corresponding observed free energy is also constant.

The fixed point strategy relies on two different ways of updating the bias (the {\em updating functions} $F_t$ and $G_t$ are increasing functions such that $G_t(0) = 0$):  
\begin{itemize}
\item The first strategy, which may be called Adaptive Biasing Potential (ABP) method, is inspired by~(\ref{eq:eq_ABP}). The bias is updated in its potential form, preferably increased (resp. decreased) for reaction coordinate values such that the observed free energy is high (resp. low):
\begin{equation}
\label{eq:ABP_update}
\text{(ABP)}\qquad
\partial_{t} A_{\text{bias}}(t,z) = 
F_{t}( A_{{\rm pot},\text{obs}}(t,z) );
\end{equation}
\item The second strategy, the usual ABF method, is inspired by~(\ref{eq:eq_ABF}). The bias is updated through the mean force: the biasing force is increased (resp. decreased) for reaction coordinate values such that the observed mean force is positive (resp. negative): 
\begin{equation}
\label{eq:ABF_update}
\text{(ABF)}\qquad
\partial_{t} A'_{\text{bias}}(t,z) = G_{t}(A'_{{\rm force}, \text{obs}}(t,z)).
\end{equation}
\end{itemize}
Let us emphasize at this point that the ABF and the ABP methods yield very different biasing dynamics, since the derivative of~(\ref{eq:ABP_calcul}) with respect to $z$ is different from~(\ref{eq:ABF_calcul}) (This would not be the case if the system was at equilibrium: the derivative of~(\ref{eq:eq_ABP}) with respect to $z$ is equal to~(\ref{eq:eq_ABF})). This difference becomes critical for multi-dimensional reaction coordinates, where the biasing force no longer derives from a potential in general.

\subsection{Consistency of the method}

Let us show that within this formalism, any stationary state of the ABP or ABF methods gives the true mean force $A'$ to be computed (and therefore the true PMF up to an additive constant). Recall at this stage that we are dealing with distribution of replicas, which arise formally in the limit of an infinite number of replicas. For a stationary state where the biasing potential has converged to $A_{\text{bias}}(\infty)$, the ergodicity property of the dynamics ensures that samples of configurations of the system are distributed according to $\psi_\infty = Z_\infty^{-1} \exp[-\beta (V-A_{{\rm bias}}(\infty,\xi))]$. 

The observed free energy or mean force given by Eqs.~(\ref{eq:ABP_calcul}) and~(\ref{eq:ABF_calcul}) then both verify  $A'_{{\rm pot},\text{obs}}(\infty,z)=A'_{{\rm force},\text{obs}}(\infty,z) = A'(z) - A'_{\text{bias}}(\infty,z)$. The updating equations Eqs.~(\ref{eq:ABP_update}) and~(\ref{eq:ABF_update}) yield respectively
\begin{equation}
\label{eq:consistance_F}
F_\infty(A(z) - A_{\text{bias}}(\infty,z)) = 0,
\end{equation}
\begin{equation} 
G_\infty(A'(z) - A'_{\text{bias}}(\infty,z)) = 0,
\end{equation}
so that (taking the derivative with respect to $z$ in~(\ref{eq:consistance_F})),
$A'_{\text{bias}}(\infty) = A'$ in both cases thanks to the strict monotonicity of the updating functions.
Let us also notice that, at convergence, the values of the reaction coordinate are distributed uniformly: $\int \psi_\infty(q) \, \delta_{\xi(q)-z} = 1$. 

However, let us emphasize that we did not give any convergence result at this point. We merely showed that, {\em if the dynamics converges}, then the limiting state is the correct one. To prove convergence starting from an arbitrary initial distribution is a difficult task, and can only be done for certain dynamics (see the corresponding results in Section~\ref{sec:application_dynamics}).

\section{Application to usual adaptive dynamics and convergence results}
\label{sec:application_dynamics}

We present in this section some applications of the formalism of Section~\ref{sec:general_framework}, and show that the usual adaptive methods can indeed be recovered. 
This is summarized in Table~\ref{tab:methods},
which gives a classification of adaptive methods.
We then give a rigorous convergence result for some class of adaptive methods.

\medskip

{\bf Table I}

\medskip

\subsection{Metadynamics}

Adaptive strategies can be used with metadynamics~\cite{LP02}. The configuration space is extended by considering an additional variable $z$ representing the reaction coordinate, and the dynamics is denoted $t\mapsto (Q_{t},Z_{t})$. The associated extended potential incorporates a coupling between this new variable and the reaction coordinate $\xi$: 
\[
V^\mu(q,z) = V(q)+ \frac{\mu}{2}(z-\xi(q))^2,
\]
for some (large) $\mu > 0$. In this case, the new reaction coordinate considered is $\xi_{\rm meta}(q,z) = z$ and the free energy is thus given by:
\[
A^\mu(z) = - \beta^{-1} \ln \int \exp(-\beta V^\mu(q,z) ) \, dq.
\] 
It is easy to check that, up to an additive constant, $A^\mu \to A$ as $\mu \to +\infty$, with $A$ given by~(\ref{eq:PMF}). The adaptive strategies presented above applied to this extended dynamics allow to recover the free energy $A^\mu$. 
The corresponding dynamics may be called meta-Adaptive Biasing Potential (m-ABP) and meta-Adaptive Biasing Force (m-ABF) methods.
   
Strategies relying on biasing potentials are reminiscent of
flooding strategies~\cite{grub95} such as the nonequilibrium
metadynamics~\cite{BLP06}. An example of an m-ABP method is metadynamics~\cite{LP02} when the biasing potential is applied to the extended variable. The updating function does not depend on time and is given by $F_t(x) = -\gamma \exp(-\beta x)$ for some constant $\gamma > 0$. The ensemble of configuration used in the adaptive update is obtained from $M$ replicas $(Q^{i,M}_t,Z^{i,M}_{t})$ running in parallel, so that 
\[
\psi_{t}(q,z) \simeq \frac1M \sum_{i=1}^M \delta_{(Q^{i,M}_t,Z^{i,M}_t)-(q,z)}. 
\]
The resulting biasing potential at time $t$ penalizes the values of the reaction coordinate already visited according to (see~(\ref{eq:ABP_update})):
\begin{eqnarray}
\label{eq:pot_metadyn}
A_{{\rm bias}}(t,z) \simeq A^M_{{\rm bias}}(t,z) = - \frac{\gamma}{M} \sum_{i=1}^M \int_0^t \delta_{Z^{i,M}_{s}- z} \, ds.
\end{eqnarray}
In the case of an overdamped Langevin dynamics with $M=1$ for example, the resulting equations of motion are therefore:
\[
\left \{ \begin{array}{cl}
dQ_t = & - \nabla V(Q_t) \, dt+ \mu(Z_t-\xi(Q_t)) \nabla \xi(Q_t) \, dt + \sqrt{2\beta^{-1}} \, dW^Q_t, \\
dZ_t = & - \mu (Z_t - \xi(Q_t)) \, dt + \sqrt{2\beta^{-1}} \, dW^Z_t - \gamma \nabla_z \left ( \int_0^t \delta_{Z_{s}- z} \, ds \right ) \, dt,\\
\end{array} \right.
\]
where the processes $W_t^Q$, $W_t^Z$ are independent standard Brownian motions.
The usual metadynamics are recovered when, as in~\cite{BLP06,RLLgMP06}, the Dirac masses $\delta_{Z_t - z}$ are discretized in the last equation and in~(\ref{eq:pot_metadyn}) using Gaussian functions. We also refer to~\cite{BLP06} for an error analysis.

\subsection{The Wang-Landau algorithm}

Another famous instance of an ABP dynamics, usually defined in discrete spaces, is the Wang-Landau~ algorithm~\cite{WL01}.  
The biasing potential is constructed in a similar fashion to~(\ref{eq:pot_metadyn}), without extending the configuration space and with only one replica. The updating function is modified during time as $F_t(x) = -\gamma(t) \exp(-\beta x)$, so that
\begin{eqnarray}
\label{eq:pot_wanglandau}
A_{{\rm bias}}(t,z) = -\int_0^t \gamma(s) \, \delta_{\xi(Q_{s})- z} \, ds.
\end{eqnarray}
If $\gamma(t) \to 0$ slowly enough, it is possible to prove the convergence of the dynamics, 
the rate of convergence of $\gamma(t)$ being controlled by the nonuniformity of the histogram of the time distribution of the reaction coordinate (see~\cite{AL04} for more precisions on the convergence results).

\subsection{The ABF method}

The usual ABF bias~\cite{HC04} is given by averaging the local force $f^{V}$ over the configurations visited by the system. It is recovered in the formalism we propose by considering one replica of the system, and an updating function of the form $G_t(x) = \gamma x$ in the limit $\gamma \to \infty$. This gives indeed:
\begin{equation} 
\label{eq:ABF_tau_zero}
A'_{\text{bias}}(t,z) = \frac{ \int f^{V}(q) \, \psi_t(q) \, \delta_{\xi(q)-z}}{\int \psi_t(q) \, \delta_{\xi(q)-z}}.
\end{equation}
Since there is only one replica, the density $\psi_t(s)$ is approximated by a trajectorial distribution, for example
\begin{equation}
\label{eq:trajectorial_distribution}
\psi_t(q) \simeq \frac1T \int_{t-T}^t \, \delta_{Q_s -q } \, ds
\end{equation}
for some averaging time $T > 0$ and $t>T$. 

\subsection{A rigorous convergence result in the ABF case}

A rigorous convergence result of the ABF algorithm with the update~(\ref{eq:ABF_tau_zero}) can be shown in the case of an overdamped Langevin dynamics~\cite{LORS}. 
It applies in the limit of an infinite number of replicas simulated in parallel using~(\ref{eq:straightforward_parallel}).
Let us detail this point. Consider the modified overdamped Langevin dynamics:
\begin{eqnarray}
\label{eq:overdamped_modified}
dQ_t = -\nabla (V +2 \beta^{-1} \ln \, |\nabla \xi| - A_{\rm bias}(t,\xi))(Q_t) \, |\nabla \xi|^{-1}(Q_t) \, dt + \sqrt{2\beta^{-1}} \, |\nabla \xi|^{-2}(Q_t) \, dW_t,
\end{eqnarray}
with the update~(\ref{eq:ABF_tau_zero}): $A'_{\rm bias}(t,z) = \langle f^V \rangle_{t,z}$.
The process $W_t$ is the standard Brownian motion.
This dynamics is the usual overdamped Langevin dynamics for the
potential ${\cal V}_t$ when $|\nabla \xi| = 1$. 
Notice that in the case of a
metadynamics-like implementation ('m-ABF'), the modified dynamics is actually
the usual overdamped Langevin dynamics since $\xi_{\rm meta}(q,z)=z$ and thus
$|\nabla \xi_{\rm meta}| = 1$.

The reason we propose to add the three terms depending on $|\nabla \xi|$ is twofold: for the dynamics~(\ref{eq:overdamped_modified})-(\ref{eq:ABF_tau_zero}), it can be proved~\cite{LORS} that (i) the distribution $\psi_t^{\xi}$ of the reaction coordinate satisfies 
\[
\partial_t \psi^{\xi}_{t} = \beta^{-1} \partial_{zz} \psi^{\xi}_{t},
\]
which is a pure diffusive behavior for the marginal law of the reaction
coordinate. In particular, the initial metastable features of the free energy landscape are overcome in a time scaling as the square of the length of the reaction coordinate set; (ii) the observed mean force $A'_{\rm bias}(t)$ converges to $A'$ when
$t \to +\infty$. The convergence is exponential, the rate of convergence
being limited by the minimum between the rate of convergence in each
submanifold ${\xi(q) = z}$ (see~\cite{LORS} for more details), and the rate of convergence in the reaction coordinate space. Let us emphasize that this proof of convergence does not assume that the system is at equilibrium at any time. 

\section{Practical implementation strategies}
\label{sec:implementation}

Relying on the definition~(\ref{eq:straightforward_parallel}) of the distribution of configurations, adaptive dynamics can be easily parallized by using a large number $M$ of replicas that interact through the biasing potential or the biasing force. 
We first show in this section how to discretize the dynamics and the biasing potential (section~\ref{sec:discretization}), and then, how this implementation can be improved using some selection process (section~\ref{sec:selection}). 

\subsection{Discretization of the biasing potential}
\label{sec:discretization}

In order to compute in practice the conditional or marginal
distributions needed to update the biasing potential, there are basically two approaches, relying either on
ergodic limits or on ensemble averages. 
Both approaches may be combined in practice in order to obtain smooth profiles. For example, when only a limited number of replicas $M$ is used, the density $\psi_t(q)$ given by~(\ref{eq:straightforward_parallel}) is not regular, and some local averaging is necessary (see {\it e.g.} Eq.~(\ref{eq:regularization}) or Eq.~(\ref{eq:regularization_autre})). 

We detail the implementation in the ABF case for example. The ABP case can be treated in a similar way (see also~\cite{RLLgMP06}).
The instantaneous conditional average of some function $h$ is typically approximated by
\[
\langle h \rangle_{t,z} \simeq \langle h \rangle^{M}_{t,z} = \frac{\sum_{i=1}^M  h(Q^{i,M}_t) \delta^\epsilon_z(\xi(Q^{i,M}_t))}{\sum_{i=1}^M \delta_z^\epsilon(\xi(Q^{i,M}_t))},
\]
where $Q^{i,M}_t$ is the $i$-th replica at time $t$ and $\delta^\epsilon_z$ is some approximation of the Dirac distribution $\delta_z$, such as a gaussian function with standard deviation~$\epsilon$ or the indicator function of an interval of size~$\epsilon$. 
In order to regularize these averages over the replicas, some time
averagings may be used (as in~(\ref{eq:trajectorial_distribution})) such as  
\begin{equation}
\label{eq:regularization}
\langle h \rangle_{t,z} \simeq \frac{\int_0^t K_\tau(t-s) \left [ \sum_{i=1}^M  h(Q^{i,M}_s) \delta^\epsilon_z(\xi(Q^{i,M}_s)) \right ] \, ds}{\int_0^t K_\tau(t-s) \left [ \sum_{i=1}^M \delta^\epsilon_z(\xi(Q^{i,M}_s)) \right ] \, ds},
\end{equation}
or 
\begin{equation}
\label{eq:regularization_autre}
\langle h \rangle_{t,z} \simeq \int_0^t K_\tau(t-s) \left [ \frac{\sum_{i=1}^M  h(Q^{i,M}_s) \delta^\epsilon_z(\xi(Q^{i,M}_s))}{\sum_{i=1}^M \delta^\epsilon_z(\xi(Q^{i,M}_s))} \right ] \, ds,
\end{equation}
with a convolution kernel $K_\tau(t)$. For instance, $K_\tau(t) = {\bf 1}_{t \geq 0} \tau^{-1} {\rm e}^{-t/\tau}$.
Many other regularizations relying on a (local) ergodicity property could of course be used. 


\subsection{Enhancing the sampling through a selection process}
\label{sec:selection}

A general strategy to improve the straightforward parallel implementation~(\ref{eq:straightforward_parallel}) is to
add a selection step to duplicate "innovating" replicas (replicas
located in regions where the sampling of the reaction coordinate is not sufficient), and kill "redundant" ones. One way to perform an efficient selection is to consider an additional
jump process quantified by a field $S(t,z)$ over the
reaction coordinate values. Each replica trajectory $(Q^{i,M}_s)$ is then weighted by
$\exp( \int_{0}^{t} S(s, \xi(Q^{i,M}_{s})) \, ds)$, which naturally gives
birth/death probabilities for the selection mechanism, in the spirit of
Sequential Monte Carlo (SMC) methods~\cite{arnaudbook} or Quantum
Monte Carlo methods (QMC)~\cite{caffarel}. A possible choice is
\begin{equation}
\label{eq:selection}
S = c \frac{\partial_{zz} \psi^{\xi}_{t} } {\psi^{\xi}_{t} },
\end{equation}
where $c$ is a positive constant. 
This method thus enhances replicas in the convex areas of the density $\psi^{\xi}_{t}$, where free energy barriers still need to be overcome. When convergence is reached, $ \psi^{\xi}_{t}$ is uniform and the selection mechanism vanishes.  

When the selection step is used with the overdamped Langevin dynamics~(\ref{eq:overdamped_modified}), it can be shown that the distribution of the reaction coordinate values $\psi^{\xi}_{t}$ still satisfies a simple diffusion equation, but with a higher
diffusion constant: 
\[
\partial_t \psi^{\xi}_{t} = (\beta^{-1} + c) \partial_{zz} \psi^{\xi}_{t}.
\]
This method thus enhances the diffusion in the reaction coordinate
space, but the convergence rate is still limited by the relaxation in each submanifold ${\xi(q) = z}$.

The jump process can be computed in practice by attaching a birth time and a death time to each replica. 
The death time is decreased when the source term $S$ given by~(\ref{eq:selection}) is positive, otherwise the birth time is decreased. When the death time is zero, the replica is replaced by another replica chosen at random. The birth process is handled in a similar way: when the birth time is zero, a replica chosen at random is replaced by the replica whose birth time is zero. We refer for example to~\cite{arnaudbook} for more precisions on the practical implementation of such a procedure, as well as other strategies to handle birth/death processes. 


\section{Numerical results}
\label{sec:numerical}

We finally present an application of the selection strategy proposed in section~\ref{sec:selection} to a model system of conformational change in solution.
We consider a two-dimensional system composed of $N$ particles in a periodic box of side
length~$l$, interacting through the purely repulsive WCA pair potential~\cite{SBB88,DBC99}: 
\[
V_{\rm WCA}(r) = 
\left \{ \begin{array}{cl}
\dps 4 \epsilon \left [ \left (\frac{\sigma}{r}\right)^{12} - \left(\frac{\sigma}{r}\right)^6 \right ] + \epsilon
& \ {\rm if \ } r \leq r_0, \\
0 & \ {\rm if \ } r > r_0.
\end{array} \right.
\]
In these expressions, $r$ denotes the distance between two particles, $\epsilon$ and $\sigma$ are two positive parameters and $r_0=2^{1/6}\sigma$.
Among these particles, two are
designated to form a solute dimer while the others are
solvent particles. For these two particles, the above WCA
potential is replaced by a double-well potential 
\[
V_{\rm S}(r) = h \left (  1 - \frac{(r-r_0-w)^2}{w^2} \right)^2,
\]
where $h$ and $w$ are two positive parameters.  
The potential $V_{\rm S}$ exhibits two energy minima, one corresponding to the compact state $r=r_0$, and one corresponding to
the stretched state $r=r_0+2w$. The energy barrier separating both states is $h$. The reaction coordinate $\xi$ is the bond length $r$ of the dimer. 

In practice, the Dirac
distribution are approximated by indicator functions of intervals of size $\Delta z = 0.05$.
The parameters used for these computations are $N=16$ particles, at particle density $\rho = N/l^2 = 0.25 \sigma^{-2}$, $\sigma = 1$, $w=0.7$, $\epsilon = 1$, $h = 20$ and $\beta = 5$. We consider $M=2000$ replicas evolving according to an overdamped Langevin dynamics,
with a time step $\Delta t = 10^{-4}$.  
The reference computation is done with $M=5000$ replicas and the reference mean force profile is obtained by averaging the profiles on the time interval~$[5,10]$. The profiles are regularized
in time by using~(\ref{eq:regularization_autre}) with
$\tau/\Delta t=100$. The initial conditions are such that the dimer bond lengths of all replicas are close to $r_0$.
We consider in the sequel the interval $[z_0,z_1] = [1.1,2.55]$ (since, with the parameters chosen here, $r_0 \simeq 1.122$, $r_0 + 2w \simeq 2.522$ and~$\Delta z = 0.05$), containing~$n=30$ bins.

\begin{figure}[h]
\includegraphics[width=17cm]{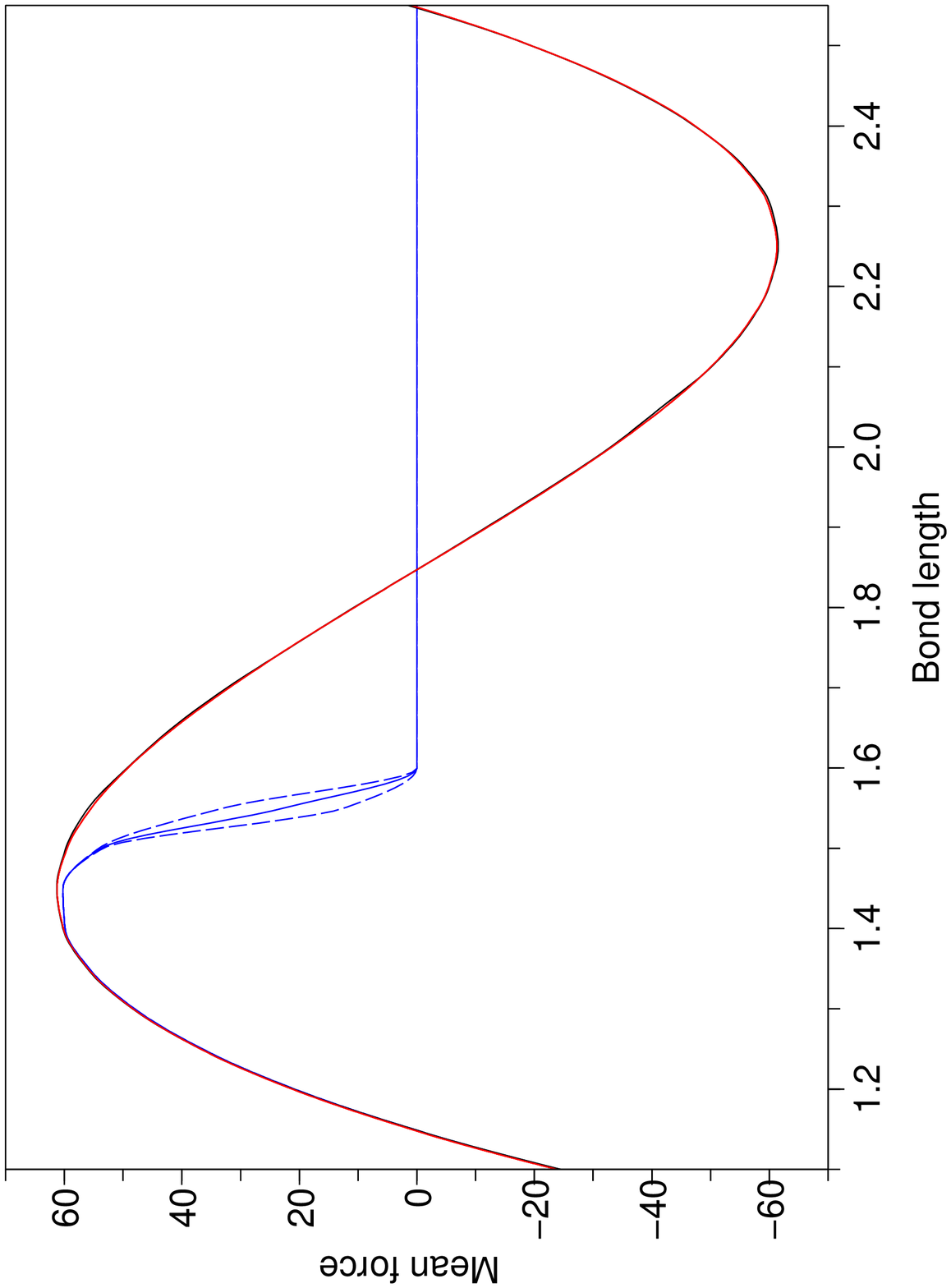}
\caption{\label{fig:resultat} Leli\`evre et al., Journal of Chemical Physics.}
\end{figure}

We present in Figure~\ref{fig:resultat} free energy difference profiles (averaged over~$K=100$ independent realizations) obtained with the parallel
ABF dynamics~(\ref{eq:ABF_tau_zero}),
with and without the birth/death selection term~(\ref{eq:selection})
(with~$c=10$), at a fixed time~$t_{\rm figure} = 0.1$. 
The standard deviation of the profiles~$(A'_1,\dots,A'_K)$ for $K$~independent realizations is
\[
\dps \sigma_{A'}(z) = \sqrt{ \frac1K \sum_{k=1}^K (A'_k(z) - {\cal A}'(z))^2},
\] 
where ${\cal A}'(z) = \frac1K \sum_{k=1}^K A'_k(z)$ is the mean force averaged over all the realizations.
The associated 95\% confidence intervals (or errors bars) are
\begin{equation}
\label{eq:confidence_interval}
[{\cal A}_-'(z), \ {\cal A}_+'(z) ] = \left [ {\cal A}'(z) - \frac{1.96}{\sqrt{K}}\sigma_{A'}(z), \ {\cal A}'(z) + \frac{1.96}{\sqrt{K}}\sigma_{A'}(z) \right ].
\end{equation}
The curves plotted in solid lines in Figure~\ref{fig:resultat} are the averages~${\cal A}'$, and the curves plotted in dashed lines are 
${\cal A}_-'$ and~${\cal A}_+'$. Notice that the mean force profile obtained when the selection process is turned on is converged (since the curves ${\cal A}'$, ${\cal A}_-'$, ${\cal A}_+'$ and the reference curve are almost indistinguishable).

\medskip

{\bf Figure I}

\medskip

The comparison shows that the selection process improves the rate of convergence of the algorithm
and accelerates the exploration process on the free energy surface. 
Indeed, the profile obtained when the selection process is turned on is quickly really close to the reference profile. On the other hand,
with a straightforward parallelization, only a small fraction of replicas has escaped from the initial free energy metastable state at time $t_{\rm figure}$ 
to explore the free energy metastable set corresponding to bond lengths around $r_0 + 2w$.

\begin{figure}[h]
\includegraphics[width=17cm]{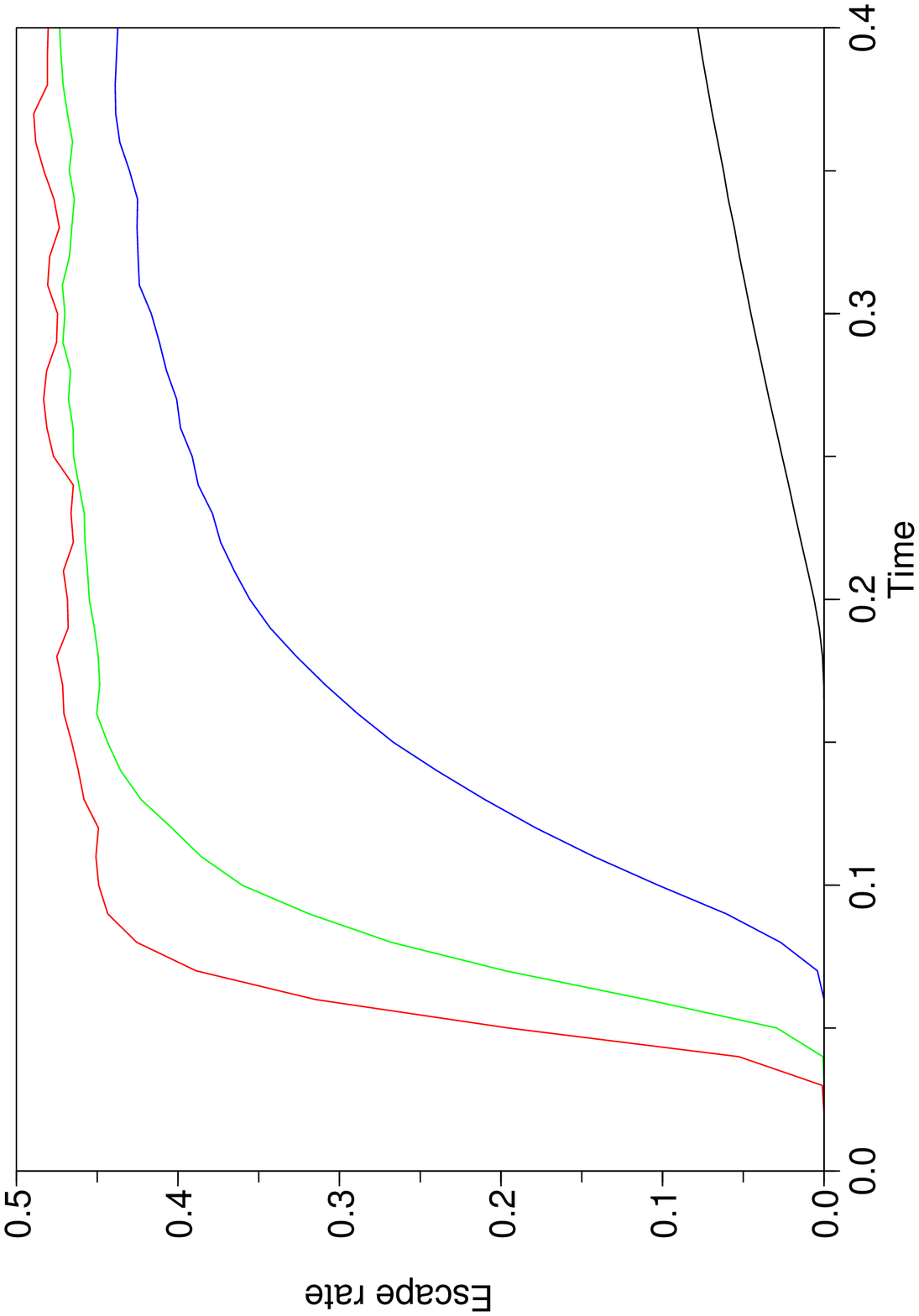}
\caption{\label{fig:escape} Leli\`evre et al., Journal of Chemical Physics.}
\end{figure}

To precise these qualitative features, we further perform two quantitative studies for several values of~$c$:
\begin{enumerate}[\quad (i)]
\item Table~\ref{tab:convergence} and Table~\ref{tab:replicas} make precise the convergence of the profiles to the reference profile in a quantitative way. The measure of error we consider is
\[
\delta A = \max_{z_0 \leq z \leq z_1} |{\cal A}(z) - A_{\rm ref}(z)|,
\] 
where $A_{\rm ref}$ is the reference profile, and ${\cal A}(z) = \int_{z_1}^z {\cal A}'$ is the averaged potential of mean force, obtained as the integral of the mean force averaged over all the realizations. In practice, we consider the following approximated deviation between PMF profiles:
\begin{equation}
\label{eq:deviation_PMF}
\delta A_n = \max_{0 \leq i \leq n} \left | \sum_{j=1}^i {\cal A}'(s_j) - A'_{\rm ref}(s_j) \right | \Delta z.
\end{equation}
A 95\% confidence interval is obtained as~$[\delta^- A_n,\delta^+ A_n]$, with
\[
\delta^\pm A_n = \max_{0 \leq i \leq n} \left | \sum_{j=1}^i {\cal A}'(s_j) \pm \frac{1.96}{\sqrt{K}} \sigma_{A'}(s_j) - A'_{\rm ref}(s_j) \right | \Delta z.
\]
\item Figure~\ref{fig:escape} presents the fraction of replicas which have crossed the free-energy barrier (averaged over the~$K=100$ realizations), {\it i.e.} the instantaneous fraction of particles such that $r \geq r_0+w$. Notice that we expect this fraction to converge to~0.5 (up to some errors due to statistical fluctuations and to the binning of~$[z_0,z_1]$). 
\end{enumerate}

\medskip

{\bf Table II}

\medskip

{\bf Table III}

\medskip

{\bf Figure II}

\medskip

As can be seen from the different escaping profiles of Figure~\ref{fig:escape}, the selection process really accelerates the transition from one free energy metastable state to the other. This is due to the fact that the birth and death jump process triggers non local moves, as opposed to the traditional diffusive exploration of adaptive dynamics. The numerical results of Table~\ref{tab:convergence} show that it is very interesting to consider a selection process, especially at the early stages of the simulation. This selection is even more efficient when the number of replicas increases (see Table~\ref{tab:replicas}). 
In conclusion, the selection process seems to be an efficient tool to improve the exploration power of the
adaptive dynamics.

\section*{Acknowledgements}

{This work was supported by the ANR INGEMOL of the French Ministry of Research. The authors acknowledge very fruitful discussions with Felix Otto and Christophe Chipot, as well as a careful rereading of the manuscript by Eric Canc\`es and Fr\'ed\'eric Legoll.}

\newpage

\section*{Table and Figures captions}

\begin{itemize}
\item {\bf Table I}. Classification of adaptive methods.
\item {\bf Figure I}. (color online) Free energy difference profiles obtained with the parallel ABF algorithm (in reduced units), for a time $t_{\rm figure}=0.1$ and averaged over~$K=100$ independent realizations: with birth/death process ($c=10$, red), without birth/death process (blue), reference computation (black). 
Solid line: average mean force; dashed lines: upper and lower bounds of the 95\% confidence intervals (see Eq.~(\ref{eq:confidence_interval})).
\item {\bf Table II}. Deviation~$\delta A_n$ from the reference PMF profile (given by Eq.~(\ref{eq:deviation_PMF})) as a function of the selection parameter~$c$ ($c=0$ when the selection is turned off) and the simulation time~$t_{\rm simu}$. The 95\% confidence interval $[\delta^- A_n, \ \delta^+ A_n]$ is given in brackets ($K=100$ realizations). 
\item {\bf Table III}. Deviation~$\delta A_n$ from the reference PMF profile (and associated error bars) when $c=10$ for different number of replicas ($K=50$ realizations).
\item {\bf Figure II}. (color online) Average fraction of the replicas in the region $r \geq r_0 + w$ as a function of time, for $c=0$ (no selection, black), $c=2$ (blue), $c=5$ (green), $c=10$ (red).
\end{itemize}

\newpage

\begin{table}
\begin{center}
\begin{ruledtabular}
\begin{tabular}{ccc}
& Adaptive Biasing & Adaptive Biasing \\
& Force ($\partial_t A_{\rm bias}'$) & Potential ($\partial_t A_{\rm bias}$) \\
\hline
\hline
Dimension $n$ ($V$) & ABF~\cite{DP01,DPW02,HC04} & ABP~\cite{WL01} \\
Dimension $n+1$ ($V^\mu$) & m-ABF & m-ABP~\cite{BLP06,RLLgMP06} \\
\hline
\end{tabular}
\end{ruledtabular}
\caption{\label{tab:methods} Leli\`evre et al., Journal of Chemical Physics.}
\end{center}
\end{table}

\vspace{15cm}

\newpage

\newpage

\begin{table}
\begin{center}
\begin{ruledtabular}
\begin{tabular}{cccccc}
c & $t_{\rm simu}=0.05$ & 0.1 & 0.2 & 0.4 \\
\hline
\hline
0 & 9.51 (7.73-11.3) & 18.0 (14.8-21.2) & 19.5 (18.3-20.7) & 0.066 (0.056-0.075) \\
2 & 20.4 (17.0-23.8) & 5.69 (5.55-5.82) & 0.020 (0.016-0.023) & 0.034 (0.029-0.038) \\
5 & 22.9 (20.9-24.9) & 0.22 (0.19-0.25) & 0.027 (0.022(0.032) & 0.026 (0.022-0.031) \\
10 & 10.4 (10.4-10.4) & 0.035 (0.029-0.041) & 0.028 (0.023-0.032) & 0.032 (0.027-0.037) \\
\hline
\end{tabular}
\end{ruledtabular}
\caption{\label{tab:convergence} Leli\`evre et al., Journal of Chemical Physics.}
\end{center}
\end{table}

\newpage

\begin{table}
\begin{center}
\begin{ruledtabular}
\begin{tabular}{cccccc}
number of replicas & $t_{\rm simu}=0.05$ & 0.1 & 0.4 \\
\hline
\hline
1000 & 23.3 (20.4-26.3) & 0.45 (0.39-0.50) & 0.064 (0.054-0.074) \\
2000 & 11.2 (11.2-11.2) & 0.034 (0.025-0.042) & 0.032 (0.024-0.039) \\
10000 & 2.05 (1.54-2.56) & 0.026 (0.019-0.033) & 0.022 (0.016-0.028) \\
\hline
\end{tabular}
\end{ruledtabular}
\caption{\label{tab:replicas} Leli\`evre et al., Journal of Chemical Physics.}
\end{center}
\end{table}

\newpage

\end{document}